\documentclass[12pt]{article}

\usepackage{graphicx}
\usepackage{amsmath}
\usepackage{amsfonts}
\usepackage{amssymb}
\usepackage{hyperref}
\usepackage{slashed}

\newcommand{\be}{\begin{equation}}
\newcommand{\ee}{\end{equation}}
\newcommand{\bea}{\begin{eqnarray}}
\newcommand{\eea}{\end{eqnarray}}
\newcommand{\ba}{\begin{array}{ll}}
\newcommand{\ea}{\end{array}}

\newcommand\eV{\text{eV}}

\newcommand\GeV{\text{GeV}}
\newcommand\TeV{\text{TeV}}

\begin{document}

\def\lsim{\mathrel{\raise.3ex\hbox{$<$\kern-.75em\lower1ex\hbox{$\sim$}}}}
\def\gsim{\mathrel{\raise.3ex\hbox{$>$\kern-.75em\lower1ex\hbox{$\sim$}}}}

\centerline{\Large Probing the Origin of Neutrino Mass: from GUT to LHC }
\vspace{5mm}
\centerline{\large Goran Senjanovi\' c}
\centerline{{\it ICTP, Trieste, Italy }}
\vspace{5mm}
\centerline{\large\sc Abstract}
\begin{quote}
\small
  
   Ever since the Majorana classic work, the nature of neutrino has been one of the central questions of the
   weak interaction physics. If neutrino is its own antiparticle, the immediate consequence is lepton number 
   violation through the neutrinoless
   double beta decay. However, colliders such as the LHC offer a hope of seeing
   directly the same phenomenon, and moreover the Majorana nature of new particles needed to complete
   the Standard Model. I review here the salient features of these phenomena, and then turn to grand unification
   as a way of probing the scale of the relevant new physics. The minimal supersymmetric SO(10) theory basically
   eliminates the LHC physics, but leads naturally to the large atmospheric mixing angle and furthermore predicts
  $\theta_{13} \simeq 10^\circ$ without any additional symmetries, in agreement with new T2K and MINOS data. It also assures that the R-parity is exact and thus leads to the stable LSP as a natural Dark Matter candidate. The minimal
  realistic SU(5) theory, on the other hand, predicts the light fermion triplet, below TeV, as a source of neutrino 
  Majorana mass. 
    
\end{quote}

 {
 \renewcommand{\thefootnote}%
   {\fnsymbol{footnote}}
 \footnotetext[4]{Based on the talk for the Neutrino Roadmap 2012, Paris, May 2011 .}
}

\tableofcontents

 \section{Introduction: from Majorana to see-saw}

We know that neutrinos are massive but light \cite{Strumia:2006db}. 
We can account for tiny neutrino masses with only the Standard Model (SM) degrees of freedom via the
effective \cite{Weinberg:1979sa} $d=5$ effective operator
\begin{eqnarray}
{\mathcal L}=Y_{ij}\frac{L_iHHL_j}{M}, 
\end{eqnarray}
where $L_i$ stands for  left-handed leptonic doublets and  H  for the usual  Higgs doublet (with a vev $v$). This
in turn produces neutrino Majorana mass matrix
\begin{eqnarray}
m_{\nu}= Y\frac{v^2}{M}. 
\label{emenu}
\end{eqnarray}

\begin{figure}
\begin{center}
\includegraphics[scale=0.8]{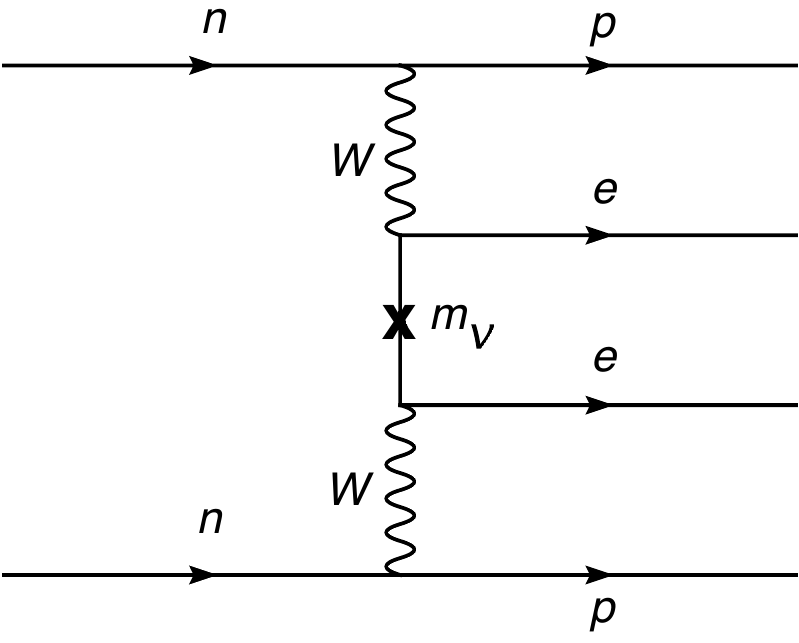} \hspace{1cm}
\includegraphics[width=5.5cm]{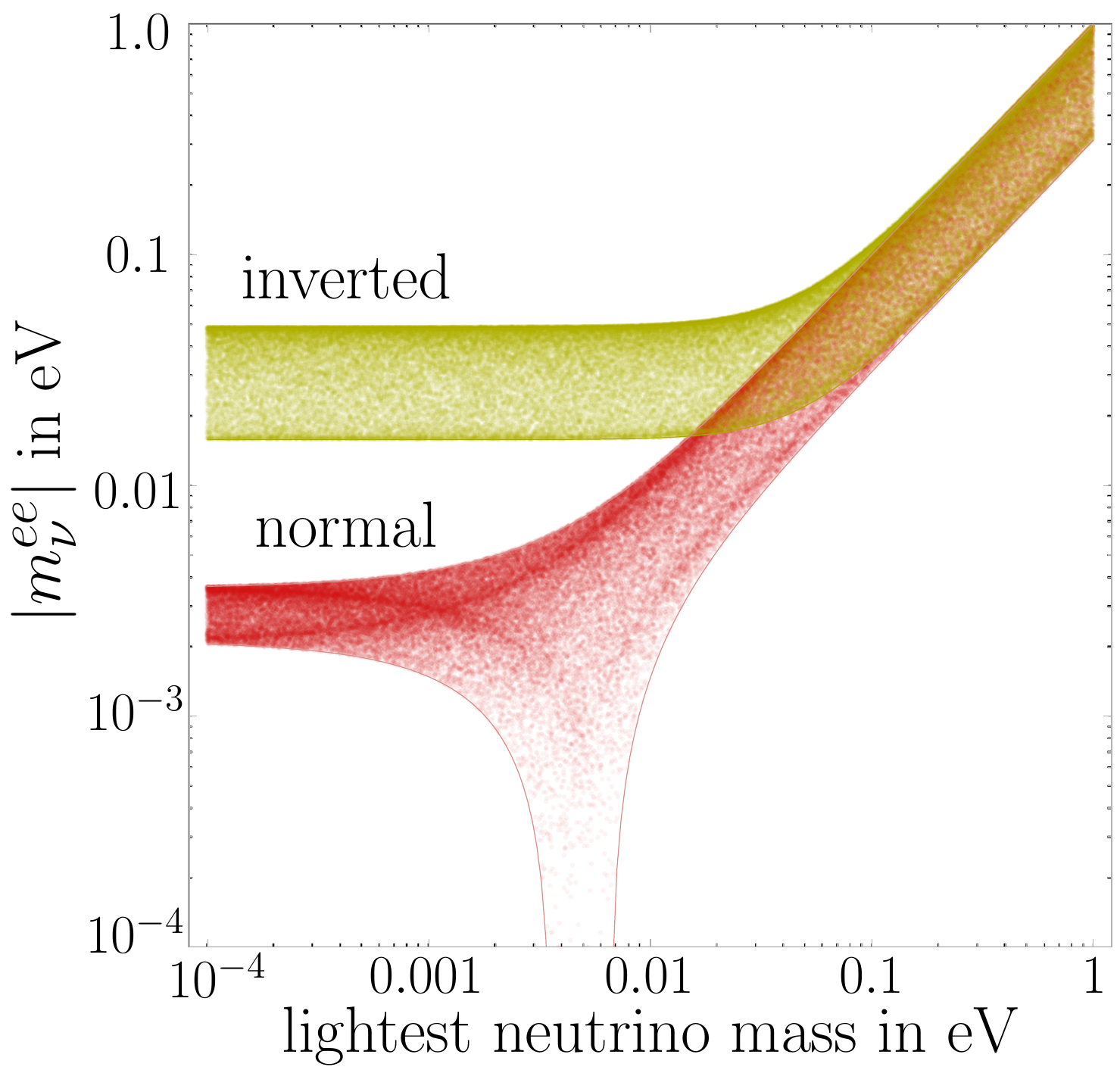}
\caption{Neutrinoless double beta decay through the neutrino Majorana mass (left), and canonical contribution from light neutrino mass (right).
          The mixing angles are fixed at $\{\theta_{12}, \theta_{23}, \theta_{13}\} = \{35^{\circ}
          ,45^{\circ},7^{\circ}\}$, while the Dirac and Majorana phases vary in the interval
          $\{0,2\pi\}$. }\label{bb0v}
\end{center}
\end{figure}

The non-renormalizable nature of the above operator signals the appearence of new physics through the mass scale $M$. The main consequence is $\Delta L=2$ violation of lepton number, which can be observed through:
\begin{itemize}
  
 \item 
   neutrinoless double beta decay ($0 \nu 2 \beta$)~\cite{Racah:1937qq}.  The canonical contribution due to neutrino Majorana mass is measured by the 1-1 element of the matrix $m_\nu$ in (\ref{emenu}). The left-hand side of Fig.\ref{bb0v}, shows the process, while in the right panel one can see the $|m_\nu^{ee}|$ element as a function of the lightest neutrino mass;

 \item 
   same sign charged lepton pairs in colliders~\cite{Keung:1983uu}.  This was the first proposal to search for lepton number violation at colliders, and is complementary to the neutrinoless double beta decay. This has only recently received wide attention, but it may be our best bet in probing directly the origin of neutrino mass.

\end{itemize}

If the scale $M$ is huge, there is no hope of direct observation of the relative new physics. It is often said that large $M$ is more natural, for then Yukawas do not have to be small.  For example, $M = 10^{13} \text{GeV}- 10^{14} \text{GeV}$ corresponds to $Y$ of order one.  However, small Yukawas are natural in a sense of being protected by chiral symmetries.

In order to get a window to new physics, we need a renormalizable theory of the above effective operator.  In the minimal scenario there are three ways of producing it, through~\cite{Senjanovic:2009at}
\begin{itemize}
\item
fermion singlets, so-called right-handed neutrinos (seesaw, or type I seesaw)~\cite{seesaw};
 \item
bosonic weak triplet (type II seesaw)~\cite{typeII}:
 
\item
 fermion weak triplet (type III seesaw)~\cite{Foot:1988aq};
 
\end{itemize}
   
%
%

%
%
%
%
%
 
 \section{Large scale see-saw: GUT}
 
 Large scale seesaw allows for Yukawas of order one. This happens naturally in 
   grand unified theories (GUT). Let us see the situation in the most popular candidates, SO(10)
   and SU(5)  GUTs.
   
 \subsection{SO(10)}

There are some features that make SO(10) special~\cite{so10}.
\begin{enumerate}
\item A family of fermions is unified in a 16-dimensional spinorial
representation ($16_F$), together with a right-handed neutrino. The
seesaw mechanism emerges naturally.
 
\item In the supersymmetric version, R-parity 
 is exact~\cite{Aulakh:1999cd}. The LSP is then stable, a natural candidate for the dark matter.

\item The unification of gauge couplings can be achieved even without
supersymmetry, with heavy right-handed neutrino and the seesaw. 

\item The minimal supersymmetric version 
gives naturally a large atmospheric mixing angle in the context of the type II
see-saw~\cite{Bajc:2002iw}.
 
\end{enumerate}

  From
\begin{equation}
16 \times 16 = 10 + 120 + \overline{126}\,,
\label{nueve}
\end{equation}
the most general Yukawa sector in general contains $10_{H}$, $120_{H}$ and $\overline{126}_{H}$.

The see-saw mechanism, whether type I or II, requires $\overline{126}$~\cite{so10}, which can be a fundamental field, or a composite of two $\overline{16_{H}}$ fields. Consider for simplicity only the former, more predictive possibility.  The minimal version contains a $10_{H}$ light Higgs and $ \overline{126}_{H}$.

\paragraph{Minimal supersymmmetric version.}

It is based on the following Higgs multiplets~\cite{Clark:1982ai}:
$10_H$, $ \overline{126}_{H}$ and $210_{H}$. It has a plethora interesting predictions, such as
naturally large leptonic and small quark mixings~\cite{Bajc:2002iw}, without any additional symmetries. It predicts~\cite{Goh:2003sy}
$\theta_{13} \simeq 10^\circ$, which is in accord with the recent T2K ~\cite{Abe:2011sj} and MINOS~\cite{minos}. These are 
remarkable results for they follow from the pure SO(10) structure, without any ad-hoc flavor symmetries 
assumed. This shows that small quark and large lepton mixing angles are nothing special, but simply a
consequence of a strongly broken Pati-Salam quark-lepton symmetry.

 Missing: predictions for proton decay branching ratios, but see a version with heavy sfermions~\cite{Bajc:2008dc}.

\paragraph{Minimal non-supersymmmetric version.}

Needs to be worked out~\cite{Bajc:2005zf}.

 \subsection{SU(5)}
 
 \paragraph{Minimal non-supersymmmetric version.}

  The minimal $SU(5)$ theory  fails for two reasons:
%
%
\begin{itemize}
\item
gauge couplings do not unify

\item
neutrinos as massless as in the SM.

\end{itemize}
A simple extension cures both problems: add an extra fermionic $24_F$~\cite{Bajc:2006ia}. 
 Under $SU(3)_C\times SU(2)_W\times U(1)_Y$:
  $24_F=(1,1)_0+(1,3)_0+(8,1)_0+(3,2)_{-5/6}+(\bar 3,2)_{5/6}$.  The unification forces the 
  triplet  fermion to lie  below TeV, with the color octet around
  $10^7$ GeV. The singlet mass is not determined, and the rest must be at the GUT scale.

     With the notation $S=(1,1)_0$ (singlet),  $T=(1,3)_0$ (triplet), it is evident that we have hybrid Type-I and Type-III seesaw,
\begin{eqnarray}
(M_\nu)^{ij}=v^2\left(
\frac{y_{T}^{i}y_{T}^{j}}{m_{T}}+
\frac{y_{S}^{i}y_{S}^{j}}{m_{S}}\right).
\end{eqnarray}
An immediate consequence is one massless neutrino, with a hierarchical spectrum.

\paragraph{Minimal supersymmmetric version.}
   
      Perfectly consistent theory~\cite{Bajc:2002pg}, including proton decay and unification constraints.
 It can even include supersymmetry breaking without a hidden sector~\cite{Bajc:2006pa}, however 
 it cannot say much about charged fermion masses due to the plethora of needed higher dimensional
 terms. Neutrino masses and mixings can follow from bilinear R-parity violating terms~\cite{Kaplan:1999ds}.

  \section{Low scale seesaw: LHC}

    In the case of low scale seesaw, one can look for collider signatures. However, this is not motivated very strongly
    since it requires taking ad-hoc some states that can serve as a source of neutrino mass, and keeping them light.
    Instead, it is preferable to study theories of neutrino masses and mixings. A natural candidate
    is provided by the left-right symmetric theories. These theories were prophetic regarding neutrino mass, and the 
    curse of its apparent largeness turned into the blessing with the advent of
     the seesaw mechanism. After going through their collider signatures, I discuss briefly the 
    resulting situation in the simplifying cases of the three types of seesaw.

\subsection{Left-Right Symmetry}
 
\begin{figure}[b!]
\begin{center}
\includegraphics[scale=0.8]{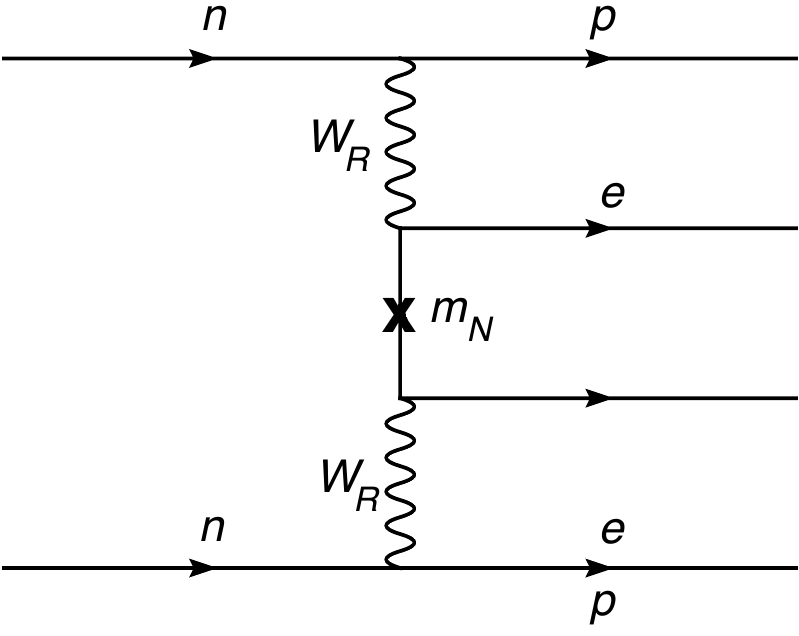} \hspace{1cm}
\includegraphics[width=5.5cm]{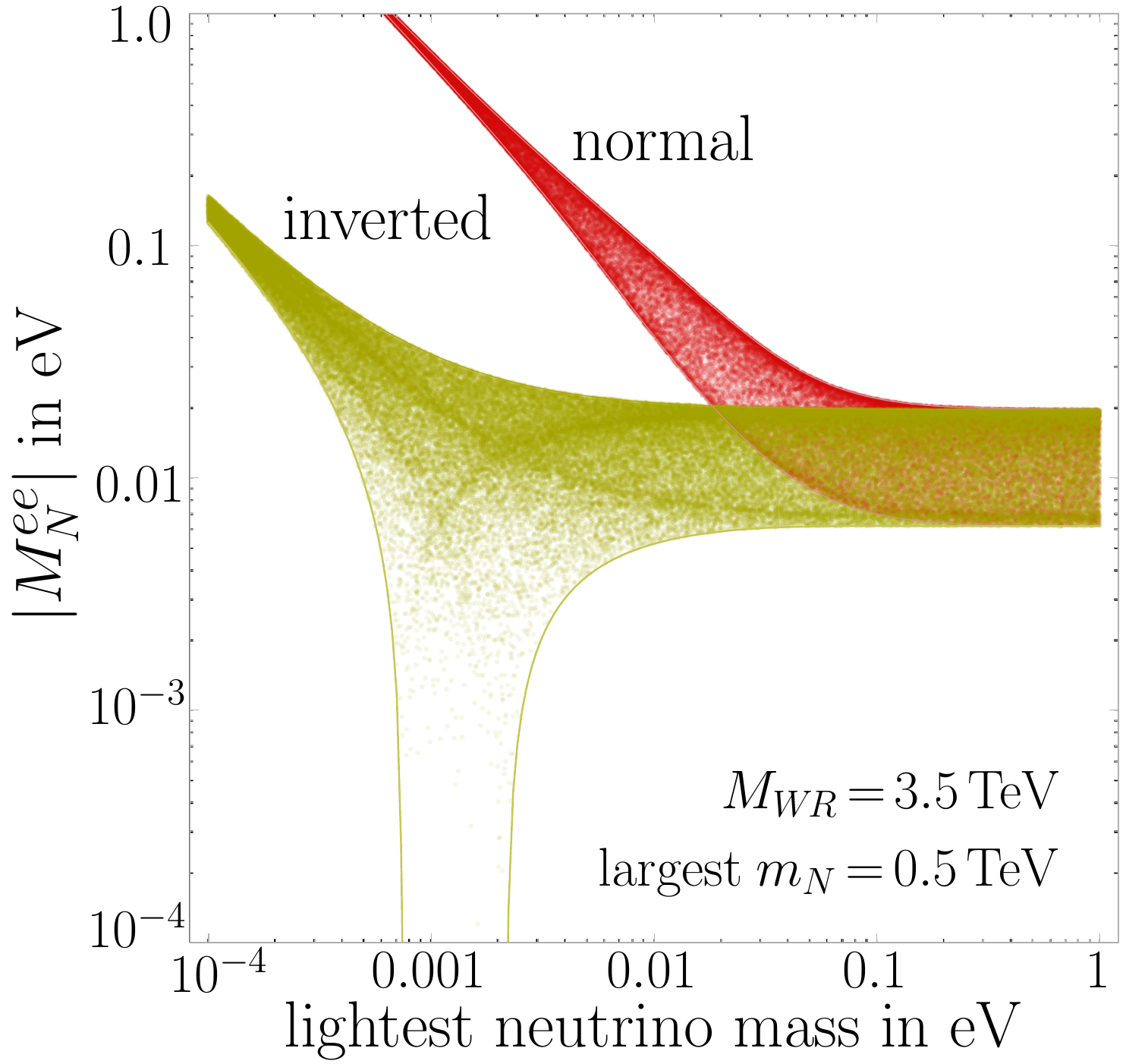}%
\caption{Neutrinoless double beta decay induced by the right-handed gauge boson and right-handed neutrino (left) and new physics contribution, taken from~\cite{Tello:2010am} (right). $M_N^{ee}$ is the LR analogue of $m_\nu^{ee}$, as explained in~\cite{Tello:2010am}.  }
\label{bb0v2}
\end{center}
\end{figure}

$L-R$ symmetric theories~\cite{so10} are based on the $SU(2)_L\times SU(2)_R \times U(1)_{B-L}$ gauge group augmented
by parity or charge conjugation.  Then:

\begin{itemize}
\item

$W_L$ implies $W_R$,

\item
$\nu_L$   implies $\nu_R$, with $m_{\nu_R}$ of order $M_R$ through the breaking of $L-R$ symmetry,

\item
Type-I seesaw:  connects neutrino  mass to the scale of parity restoration.

\end{itemize}
These facts lead immediately to the new contribution~\cite{Mohapatra:1980yp} to the neutrino-less double beta decay mentioned above, see Fig.~\ref{bb0v2}.  With $W_R$ in the TeV region and the right-handed neutrino mass $m_N$ in the 100 GeV -TeV region, this contribution can easily dominate over the left-handed one. Neutrino mass can even go to zero (vanishing Dirac Yukawa) while keeping the $W_R$ contribution finite. This was revisited and studied carefully in the context of the Type-II seesaw~\cite{Tello:2010am}.

The new physics source of the neutrinoless double beta decay, may not be just an amusing possibility but even a must. The point is that cosmology keeps pushing down the sum of neutrino masses. The recent study~\cite{Hannestad:2010yi} concludes $\Sigma m_\nu \lesssim 0.44$ \eV (@ 95 \% CL), while ~\cite{Seljak:2006bg} finds an even stronger limit $\Sigma m_\nu \lesssim 0.17$ \eV (@ 95 \% CL).

The observation of neutrinoless double beta decay in the present and near future experiments, such as GERDA and CUORE~\cite{Racah:1937qq}, may thus require new physics, especially if the sum of neutrino masses is to go down even more. As it is, there is already a mild tension between cosmological limits and the claimed positive result~\cite{KlapdorKleingrothaus:2004wj}.
 
The discrete $L-R$ symmetry can be P or C. From $K_L-K_S$ mass difference one gets the limit $M_{W_R}\gtrsim4\,\TeV$ in the case of P, whereas for C, $M_{W_R}\gtrsim2.5\,\TeV$~\cite{Maiezza:2010ic}.  

\begin{figure}
\begin{center}
\includegraphics[width=6cm]{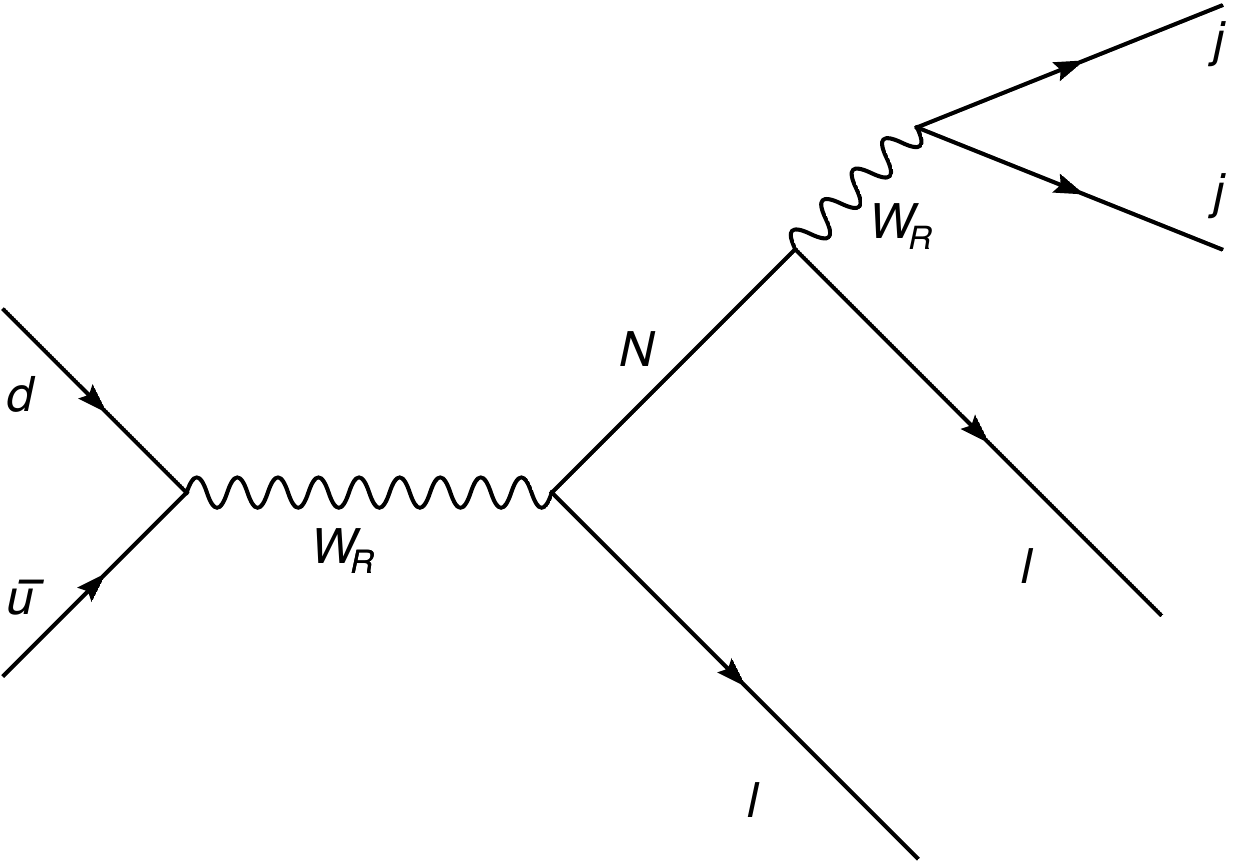} \hspace{1cm}
\includegraphics[width=7.5cm]{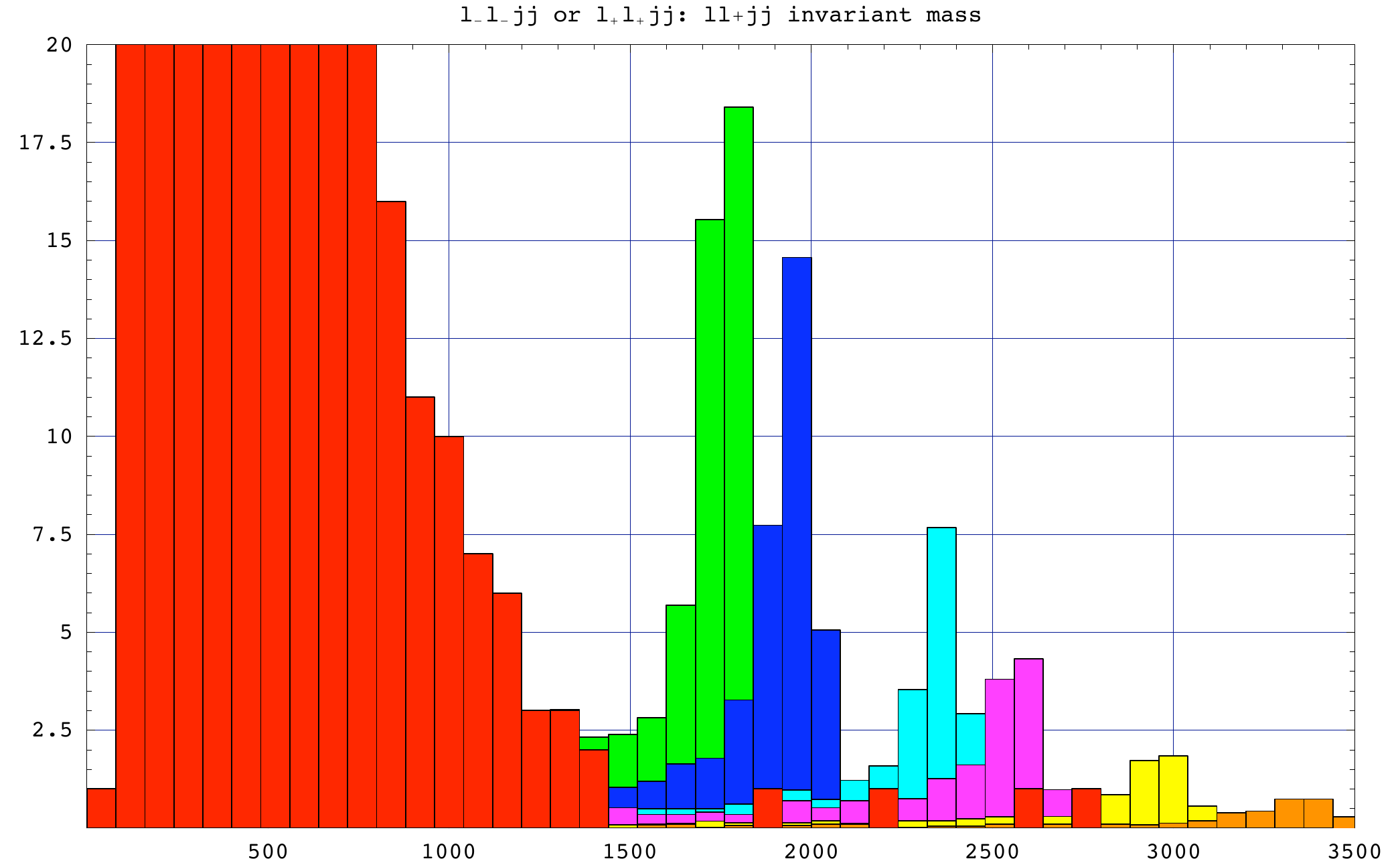} 
\caption{The production of $W_R$ and the subsequent decay into same sign leptons and two jets through
the Majorana character of the right-handed neutrino (left) and number of events (right) as a function of energy (GeV) for ${\rm L}=10{\rm fb} ^{-1}$ (courtesy of F. Nesti)
 where ${\rm M_R}$ (TeV) is  taken to be: $1.8;\,2.0;\,2.4;\,2.6;\,3,0;\,3.4$.}
\label{wr}
\end{center}
\end{figure}

\paragraph{Colliders.} Produce $W_R$ through Drell-Yan as in Fig. \ref{wr}.
The right-handed gauge boson decays into a right-handed neutrino and a charged lepton.  The 
right-handed neutrino, being a Majorana particle, decays equally often into charged leptons or anti-leptons~\cite{Keung:1983uu}
 and jets. This offers the only possible way of probing directly the Majorana nature
of any particle.

 In turn one has exciting events of same sign lepton pairs and two jets, as a clear signature of lepton number violation.
This is the collider analog of neutrino-less double beta decay, and it allows for the determination of $W_R$ mass as
shown in the right side of  Fig. \ref{wr}, offering
\begin{itemize}
\item   direct test of parity restoration through a  discovery of $W_R$,
\item direct test of lepton number violation through a Majorana nature of $\nu_R$,
\item  determination of $W_R$  and $N$ masses.
\end{itemize}
A detailed study \cite{Ferrari:2000sp} concludes a possible probe at the LHC of $W_R$ up to 4 \TeV\ and $\nu_R$ between $100 - 1000$  \GeV\ for integrated luminosity of 30 ${\rm fb} ^{-1}$, and $W_R$ up to 5.5 \TeV\ for a luminosity of 300 ${\rm fb} ^{-1}$ The flavor structure of this process would help determine (at least partially) the right-handed lepton mixing matrix, which in turn would allow to make predictions for flavor violating processes (LFV) and the neutrinoless double beta decay~\cite{Tello:2010am}. It is impressive that 
the early LHC data already established a limit $M_{W_R}\gtrsim1.4\,\TeV$~\cite{Nemevsek:2011hz} for a wide
range of right-handed neutrino mass.
   
   It is worth noting that the same signatures can be studied in the SM with $\nu_R$~\cite{Senjanovic:2009at}, but it 
   requires miraculous cancellations of large Dirac Yukawa couplings in order to keep neutrino masses small.
  When a protection symmetry is called for, one ends up effectively with lepton number conservation and the 
  phenomenon disappears \cite{Kersten:2007vk}.

The $L-R$ theory possesses naturally~\cite{Mohapatra:1980yp} also Type-II seesaw. The Type-II offers another potentially interesting signature: pair production of doubly charged scalars which decay into same sign lepton (anti lepton) pairs. This can serve as a determination of the neutrino mass matrix in the case when Type-I is not present or very small~\cite{Kadastik:2007yd}.

\subsection{More SU(5)}    
    
     We saw that unification predicts the mass of the fermion triplet below TeV, and thus it becomes accessible to the colliders such as Tevatron and LHC.
 It can be produced through gauge interactions (Drell-Yan)
\begin{eqnarray}
pp\to W^\pm +X\to T^\pm T^0 +X\nonumber\\
pp\to (Z\,{\rm or}\, \gamma)+X\to T^+T^-+X\,.\nonumber
\end{eqnarray}
%

The best channel is like-sign dileptons + jets
\begin{eqnarray}
BR(T^\pm T^0\to l_i^\pm l_j^\pm +4{\rm jets})\approx\frac{1}{20}\times
\frac{|y_T^i|^2|y_T^j|^2}{(\sum_k|y_T^k|^2)^2}\nonumber
\end{eqnarray}

Same couplings $y_T^i$ contribute to $\nu$ mass matrix and $T$ decays, so that T decays can serve to probe the
neutrino mass matrix~\cite{Bajc:2006ia};\cite{Arhrib:2009mz}.
 
 With an integrated luminosity of 10 (100)\,${\rm fb} ^{-1}$ one could find the fermionic triplet $T$ at 14 TeV LHC for $M_T$ up to about 450 (700) \GeV~\cite{Arhrib:2009mz}.

\section{Summary and Outlook }

I discussed here an  experimental probe of Majorana neutrino mass origin, both at colliders 
and through the neutrinoless double beta decay.   
It is shown that a TeV scale $L-R$ symmetry would have spectacular signatures at LHC, 
with a possible discovery of $W_R$ and $\nu_R$.  Moreover, if neutrinoless double beta decay were 
to be established, and neutrino masses were to be pushed down by cosmology, the TeV L-R scale
could be a must. It is gratifying that these fundamental experiments are going on at the same time.

A case is made for a predictive grand unified theory: 
minimal $SU(5)$ with extra fermionic adjoint.   A weak fermionic triplet must lie in the TeV range, with good chances for discovery at LHC. Its decays probe directly the masses and mixings.
 
  I also discussed the $SO(10$) grand unified theory of neutrino masses and mixings. The minimal supersymmetric version succeeds in connecting
  large leptonic and small quark mixing angles, and predicts the 13 leptonic mixing in accordance with the new 
  T2K and MINOS results.  It is worth considering the implications of this results. It is often argued that the quark and 
  lepton mixing angles ought to be similar and that their disparity is a deep question. However, this question 
  makes sense only in a theory with quark-lepton symmetry such as SO(10), and yet, a minimal such theory
  gives automatically this dichotomy once the quark-lepton symmetry is broken at the large GUT scale. Moreover,
  it also seem to happen in the SO(10) theory with the radiative seesaw mechanism~\cite{Bajc:2005aq}.
    This theory also gives naturally a dark matter candidate in the form of the LSP since it predicts exact 
  R-parity, otherwise assumed ad-hoc in the MSSM.  This too happens automatically, without any ad-hoc
  discrete symmetries. Simply, matter parity, equivalent to R-parity is a gauge symmetry, and it has to remain
  exact or otherwise there would a light pseudo-majoron coupled to the Z-boson and thus ruled out.
 
     \section{Acknowledgements}
     
      I am grateful to Alejandra Melfo and Fabrizio Nesti for their help in preparing this report, and Miha
      Nemev\v{s}ek for a careful reading of the manuscript. I wish to thank Silvia Pascoli, Vittorio Paladino
      and other organizers of the Neutrino Roadmap 2012 for the invitation to  present an overview of the
     origin of neutrino mass.

\end{document}